\documentclass[twocolumn,showpacs,preprintnumbers,prd,superscriptaddress,nofootinbib]{revtex4-1}

\usepackage[dvips]{color}
\usepackage{epsfig}
\usepackage{amsmath}
\usepackage{graphicx}

\begin{document}

\title{Cosmic acceleration in non-flat $f(T)$ cosmology}

\author{Salvatore Capozziello}
\email{capozzie@na.infn.it}
\affiliation{Dipartimento di Fisica, Universit\`a di Napoli  ``Federico II'', Via Cinthia, I-80126, Napoli, Italy.}
\affiliation{Istituto Nazionale di Fisica Nucleare (INFN), Sez. di Napoli, Via Cinthia 9, I-80126 Napoli, Italy.}
\affiliation{Gran Sasso Science Institute, Via F. Crispi 7, I-67100, L' Aquila, Italy.}

\author{Orlando Luongo}	
\email{luongo@lnf.infn.it}
\affiliation{Istituto Nazionale di Fisica Nucleare (INFN), Laboratori Nazionali di Frascati, Via E. Fermi 40, 00044 Frascati, Italy.}
\affiliation{School of Science and Technology, University of Camerino, I-62032, Camerino, Italy.}
\affiliation{Instituto de Ciencias Nucleares, Universidad Nacional Aut́onoma de Ḿexico, AP 70543, Mexico, DF 04510, Mexico.}

\author{Richard Pincak}
\email{pincak@saske.sk}
\affiliation{Institute of Experimental Physics, Slovak Academy of Sciences Watsonova 47, SK-043 53 Košice, Slovakia,}
\affiliation{Bogoliubov Laboratory of Theoretical Physics, Joint Institute for Nuclear Research 141980 Dubna, Moscow Region, Russia.}

\author{Arvin Ravanpak}
\email{a.ravanpak@vru.ac.ir}
\affiliation{Department of Physics, Vali-e-Asr University, Rafsanjan, Iran}

\date{\today}

\begin{abstract}
We study   $f(T)$ cosmological models inserting a non-vanishing spatial curvature and discuss its  consequences on cosmological dynamics. To figure this out,  a polynomial $f(T)$ model and a  double torsion model are considered. We first analyze those models with cosmic data, employing the recent surveys of Union 2.1, baryonic acoustic oscillation and cosmic microwave background measurements. We then emphasize that the two popular $f(T)$ models enable the crossing of the phantom divide line due to dark torsion. Afterwards, we compute numerical bounds up to 3-$\sigma$ confidence level, emphasizing the fact that $\Omega_{k0}$ turns out to be non-compatible with zero at least at 1$\sigma$. Moreover, we underline that, even increasing the accuracy, one cannot remove the degeneracy between our models and the $\Lambda$CDM paradigm. So that, we show that our treatments contain the concordance paradigm and we analyze the equation of state behaviors at different redshift domains. We also take into account gamma ray bursts and we describe the evolution of both the $f(T)$ models with high redshift data. We calibrate the gamma ray burst measurements through small redshift surveys of data and we thus compare the main differences between non-flat and flat $f(T)$ cosmology at different redshift ranges. We finally match the corresponding outcomes with small redshift bounds provided by cosmography. To do so, we analyze the deceleration parameters and their variations, proportional to the jerk term. Even though the two models well fit late-time data, we notice that the polynomial $f(T)$ approach provides an effective de-Sitter phase, whereas the second $f(T)$ framework shows analogous results compared with the $\Lambda$CDM predictions.
\end{abstract}


\maketitle


\section{Introduction}

Consolidate large-scale observations indicate a phase of cosmological acceleration  \cite{Riess,Perlmutter,Spergel,Spergel2,Tegmark,Eisenstein} which occurs at late-times. The concordance paradigm cannot address this evidence with pressureless matter only since matter bids to standard gravitational attraction and is unable to speed up the universe today.

Hence, several attempts have been proposed to describe the cosmic accelerated scenario \cite{od1,od2,report,cai}. Among all, the basic idea aims to include into Einstein's energy momentum tensor a new ingredient, dubbed \emph{dark energy}, typically under the form of perfect fluid. Dark energy counterbalances the action of gravity providing an effective  negative pressure. This fluid acts to push up the universe today after a precise redshift domain, named the \emph{transition redshift}\footnote{The presence of matter and dark energy manifests a redshift at which dark energy starts dominating over matter, i.e. the transition redshift $z_{tr}$.} \cite{transizione}. The simplest and widest accepted dark energy explanation includes the cosmological constant, hereafter $\Lambda$, whose equation of state (EoS) reads: $\omega=-1$, whose origin is associated to quantum vacuum energy. The cosmological constant represents the basic ingredient behind the cosmological concordance paradigm, i.e. the $\Lambda$CDM model \cite{ab1,ab2}. Although appealing and widely accepted, the concordance paradigm is jeopardized by several shortcomings which suggest $\omega$ to evolve in terms of the redshift $z$, instead of being a pure constant.

\noindent Consequently, over past decades numerous dark energy models with time dependent EoS parameters have been discussed in the literature. The simplest one includes a scalar field as dark energy model \cite{Caldwell,Caldwell2,Armendariz,Padmanabhan,Sen,Feng,eli,Cognola:2007zu, Cognola:2006eg,Kamenshchik,Bento} which provides a constant EoS, albeit different from $\omega=-1$. Other approaches span from quantum holography, Cardassian up to parameterized dark energy, non-perfect fluids and so forth \cite{Cohen,Li,Wei2,Wei}. More recently, modifications of Einstein's gravity departing from Einstein's general relativity have reached much consensus \cite{Gao,Felice,Mota,Farajollahi,Zuntz,Camera}. This class of models handles cosmic speed up by means of prime principle, invoking that gravity breaks down into a more complicated paradigm at certain energy regimes. Those theories have been motivated even at the level of quantum gravity and refer to classes of models such as $f(R)$, $f(T)$, $f(R,G)$  and so on. Among $f(R)$ models, there exist those which have been verified by several kinds of observational and theoretical constraints. These scenarios may exhibit universe acceleration at late times and even phantom crossing \cite{noj1,noj2,noj3,noj4,noj5}.

\noindent In addition, the class of models based on modified teleparallel gravity were presented as alternative to explain inflationary phases \cite{bengo}. They provide a cosmological implication which pushes up the universe through an analytic torsion function, written as $f(T)$. This turns out to be a modification of teleparallel equivalent of general relativity (TEGR) Lagrangian \cite{Einstein,Hayashi,Einstein2}. In these treatments, \emph{dark torsion} becomes responsible for the observed speed up. In such a picture, the field equations are framed by second order differential equations in strict analogy to general relativity. This property represents a great advantage than $f(R)$ models\footnote{In the framework of metric formalism} and candidates as a viable alternative to curvature.

\noindent Recently, great attention has been devoted to investigate the main consequences of such modified theories \cite{Rafael,Puxun,Ravanpak,Rong,Hao,Shamaila,Miao,Yi,Baojiu,Rafael2,Yi2,Kazuharu,Rui,James,Gabriel,Ferraro3,Ulhoa,Nashed,Sharif,Lucas,Ferraro2,Poplawski,Wu,Wu1,Linder,Bamba,Ao,Bengochea,Ferraro,Wu2,Yang2} if spatial curvature does not vanish \cite{bengo}. The motivation behind the choice of non-flat cosmology is that, as usually believed, an early inflationary phase leads today to almost flat universe, albeit not exactly with a perfectly zero spatial curvature. This is not necessary if the number of e-foldings is not very large \cite{Huang}. It is still possible that there is a contribution to the Friedmann equations from the spatial curvature when studying late-time universe, though much smaller than other energy components according to observations.

Thus, assuming a non-flat universe turns out to be not only of academic interest but provides a universe which allows inflation in agreement with current cosmic puzzle \cite{infla}.

\noindent Further, gamma ray bursts (GRBs) have recently attracted much attention as possible objects to extend Hubble's diagram to very high redshifts. To this aim, the luminosity (or isotropic emitted energy) of a GRB at redshift $z$ must be evaluated from a correlation with a distance independent quantity, so that one can solve the luminosity distance $D_L(z)$, getting the distance modulus $\mu(z)$. Averaging over five different two correlation parameters and using a fiducial cosmological model to calibrate them in \cite{Schaefer} the author has compiled a sample of 69 GRBs with measured $\mu(z)$ which has been widely used to constrain cosmological parameters.

In \cite{Cardone} the authors updated the aforementioned sample, upgrading many aspects. First, they added consequences got from a recent correlation for X-ray afterglows. They even propose the use of Bayesian's inspired-fitting method to calibrate the different GRBs correlations. The mechanism provides an averaging over six correlations, which ended with the byproduct of new GRB Hubble diagram. The diagram comprises 83 objects, which has been produced with the fiducial standard model\footnote{For the sake of clearness, to avoid the \emph{a priori} chosen cosmological model, one can calibrate on a model-independent local regression estimate of $\mu(z)$ using Union supernova sample. This leads to a GRB Hubble diagram made out of 69 GRBs \cite{gaumma}.}.

In this manuscript, we check the goodness of two widely appreciated $f(T)$ models, previously discussed in \cite{Wu1}, which have reached great attention either for their simplicity or for their capability in describing late-time cosmology. We consider the extensions of such models by adding non-flat spatial curvature, in order to check whether in the framework of $f(T)$ cosmology the net effects of spatial curvature is relevant. We check when the cosmic acceleration starts and how constraints over free coefficients are modified by the presence of spatial curvature. We investigate the phantom crossing divide by fitting the models with recent observational data. We involve supernova data, baryonic acoustic oscillation and cosmic microwave background surveys. We also introduce GRBs, using in particular different sets of available data, requiring the techniques reported above. Finally, we check the matching between cosmography of $f(T)$ gravity with the numerical outcomes got from the above scheme. We find a good agreement with cosmic measurements, although error bars, evaluated up to the 3$\sigma$ cannot exclude the concordance model. Even at the level of cosmography we get suitable results, compatible with theoretical predictions and experimental bounds. In particular, cosmography seems to select the polynomial approach to $f(T)$ as the most viable candidate to reproduce an effective torsion dark energy at late-times.

The paper is structured as follows. In Sec. II, we show how to extend teleparallel gravity to $f(T)$ paradigms. In Sec. III, we propose $f(T)$ solutions at the level of cosmology, adding the spatial curvature term. In the same section, we reported the two cases of interest, proposing the polynomial $f(T)$ gravity and a more complicated example. We then analyze the most relevant properties and all basic demands associated to the models themselves. In Sec. IV, we show in detail how to handle cosmic data, developing supernovae, baryonic acoustic oscillation and cosmic microwave background radiation. We continue our discussion on cosmic data and experimental applications, adding in Sec. V the use of GRBs. In Sec. VI, we compare the so-obtained results with cosmography and finally in Sec. VII, we discuss final outlooks and perspectives of our work.


\section{Extending the Teleparalleling cosmology: general considerations and cosmological applications}

Teleparallel theories take into account vierbein fields, i.e. $e_i(x^\mu)$, $i = 0, 1, 2, 3$, which represent dynamical objects as orthonormal basis for the tangent space at each point $x^\mu$ of the manifold: $e_i \cdot e_j=\eta_{ij}$, where $\eta_{ij}=diag(1,-1,-1,-1)$. Each vector $e_i$ is described through its components $e^\mu_i$, $\mu=0,1,2,3$ in a coordinate basis, i.e. $e_i=e^\mu_i\partial_\mu$\footnote{Hereafter, Latin indexes refer to as the tangent space,
whereas Greek indexes label as manifold coordinates.}. The metric tensor can be thus obtained from the dual vierbein as $g_{\mu\nu}(x)=\eta_{ij} e^i_\mu(x)e^j_\nu(x)$. Differently from general relativity, which makes use of the torsionless Levi-Civita connection, in Teleparallel gravity one uses the curvatureless Weitzenb\"{o}ck connection, whose non-null torsion is
\begin{equation}\label{torsion}
    T^\lambda_{\mu\nu}=\hat{\Gamma}^\lambda_{\nu\mu}-\hat{\Gamma}^\lambda_{\mu\nu}=e^\lambda_i(\partial_\mu e^i_\nu - \partial_\nu e^i_\mu)\,.
\end{equation}

\noindent This tensor encompasses all the information about the gravitational field. This approach represents an alternative view to curvature and in principle can be considered perfectly equivalent to general relativity. If one extends the above formalism, it is possible to introduce the TEGR Lagrangian, which is built up through an arbitrary function of the torsion (\ref{torsion}) itself.

\noindent The teleparallel Lagrangian takes the form:
\begin{equation}\label{lagrangian}
    T={S_\rho}^{\mu\nu}{T^\rho}_{\mu\nu}\,,
\end{equation}
where
\begin{equation}\label{s}
    {S_\rho}^{\mu\nu}=\frac{1}{2}({K^{\mu\nu}}_\rho+\delta^\mu_\rho {T^{\theta\nu}}_\theta-\delta^\nu_\rho {T^{\theta\mu}}_\theta)\,,
\end{equation}
and ${K^{\mu\nu}}_\rho$ is the contorsion tensor, defined as:
\begin{equation}\label{contorsion}
    {K^{\mu\nu}}_\rho=-\frac{1}{2}({T^{\mu\nu}}_\rho-{T^{\nu\mu}}_\rho-{T_\rho}^{\mu\nu})\,,
\end{equation}
which equals the difference between Weitzenb\"{o}ck and Levi-Civita connections.

\noindent In this work, we adopt the standard formalism in which the gravitational field is driven by a Lagrangian density, as sum over $T$ and $f(T)$. In particular, the action reads
\begin{equation}\label{action}
    I = \frac{1}{16\pi G}\int{d^4xe(T+f(T))}\,,
\end{equation}
where $e=det(e^i_\mu)=\sqrt{-g}$. The action with $T$ only corresponds to TEGR. If matter couples to the metric in the standard form then action's variation with respect to the vierbein leads to \cite{Ferraro}:
\begin{eqnarray}
&& e^{-1}\partial_\mu(e{S_i}^{\mu\nu})(1+f^{'}(T))-e_i^\lambda {T^\rho}_{\mu\lambda}{S_\rho}^{\nu\mu}(1+f^{'}(T))+\nonumber \\& &
{S_i}^{\mu\nu}\partial_\mu(T)f^{''}(T)-\frac{1}{4}e^\nu_i(T+f(T))=4\pi G{e_i}^\rho {T_\rho}^\nu\,, \label{equations}
\end{eqnarray}
where primes denote differentiations with respect to $T$, ${S_i}^{\mu\nu}={e_i}^\rho {S_\rho}^{\mu\nu}$ and $T_{\mu\nu}$ is the matter energy-momentum tensor.


\section{Cosmological solutions and applications to the observable universe}

We now want to include spatial curvature into the aforementioned treatment. Thus, we may assume a non-flat homogeneous and isotropic FRW universe, in agreement with the cosmological principle. Hence, following  \cite{Ferraro4} and \cite{Ferraro5}, we have:
\begin{eqnarray}\label{metric}
    e^0=dt\,\,\,, e^1=aE^1\,\,\,, e^2=aE^2\,\,\,, e^3=aE^3\,,
\end{eqnarray}
in which $a=a(t)$ is the cosmological scale factor and $E^1$, $E^2$ and $E^3$ are expressed, for closed universe, as
\begin{eqnarray}\label{tc}
  E^1 &=& -\cos(\theta)d\psi + \sin(\psi)\sin(\theta)\times\nonumber\\
&&(\cos(\psi)d\theta - \sin(\psi)\sin(\theta)d\phi)\,,\nonumber \\
  E^2 &=&  \sin(\theta)\cos(\phi)d\psi - \sin(\psi)\times\nonumber\\
  &&[(\sin(\psi)\sin(\phi) - \cos(\psi)\cos(\theta)\cos(\phi)) d\theta \nonumber \\ &+& (\cos(\psi)\sin(\phi) + \sin(\psi)\cos(\theta)\cos(\phi)) \sin(\theta) d\phi]\,,\nonumber\\
  E^3 &=&  -\sin(\theta)\sin(\phi)d\psi - \sin(\psi)\times\\
  &&[(\sin(\psi)\cos(\phi) + \cos(\psi)\cos(\theta)\sin(\phi)) d\theta \nonumber \\ &+& (\cos(\psi)\cos(\phi) - \sin(\psi)\cos(\theta)\sin(\phi)) \sin(\theta) d\phi]\,,\nonumber
\end{eqnarray}
while for open universe as
\begin{eqnarray}\label{to}
  E^1 &=& \cos(\theta)d\psi + \sinh(\psi)\sin(\theta)\times\nonumber\\
&&(-\cosh(\psi)d\theta +i \sinh(\psi)\sin(\theta)d\phi)\,,\nonumber \\
  E^2 &=&  -\sin(\theta)\cos(\phi)d\psi + \sinh(\psi)\times\nonumber\\
  &&[(i \sinh(\psi)\sin(\phi) - \cosh(\psi)\cos(\theta)\cos(\phi)) d\theta \nonumber \\ &+& (\cosh(\psi)\sin(\phi) + i \sinh(\psi)\cos(\theta)\cos(\phi)) \sin(\theta) d\phi]\,,\nonumber\\
  E^3 &=&  \sin(\theta)\sin(\phi)d\psi + \sinh(\psi)\times\\
  &&[(i \sinh(\psi)\cos(\phi) + \cosh(\psi)\cos(\theta)\sin(\phi)) d\theta \nonumber \\ &+& (\cosh(\psi)\cos(\phi) - i \sinh(\psi)\cos(\theta)\sin(\phi)) \sin(\theta) d\phi]\,.\nonumber
\end{eqnarray}

\noindent Considering Eqs. (\ref{torsion}), (\ref{lagrangian}), (\ref{s}) and (\ref{contorsion}), we thus obtain the Lagrangian for non-flat $f(T)$ models, i.e. either closed or open universes, through the requirements of (\ref{tc}) and (\ref{to}), respectively as
\begin{equation}\label{lt}
    T=-6H^2+\frac{6k}{a^2},
\end{equation}
where $H\equiv\frac{\dot a}{a}$ is the Hubble parameter, with $k=\pm1$ for closed and open universes respectively. The substitution of the vierbein (\ref{metric}) in Eqs. (\ref{equations})
for $i=0=\nu$ yields
\begin{equation}\label{friedmann}
    6H^2+12H^2f^{'}(T)+f(T)+\frac{6k}{a^2}=16\pi G\rho\,.
\end{equation}
Besides, the equation $i=1=\nu$ is
\begin{eqnarray}\label{acceleration}
    &&48H^2f^{''}(T)[\dot{H}+\frac{k}{a^2}]-4f^{'}(T)[3H^2+\dot{H}- \frac{k}{a^2}]\nonumber\\  &&-f(T)-4\dot{H}-6H^2-\frac{2k}{a^2}=16\pi G P\,.\nonumber\\
\end{eqnarray}

Looking at Eqs. (\ref{friedmann}) and (\ref{acceleration}), it is evident that $\rho$ and $p$ are the dark sector energy density and pressure respectively for a source built up by perfect fluids only. Combining both the Friedmann equations, one accomplishes the conservation equation
\begin{equation}\label{conservation}
    \dot{\rho}+3H(\rho+P)=0\,.
\end{equation}
Here, we assume that the matter component is only made by pressureless cold dark matter and baryons, thus $\rho = \rho_m$ and $P = P_m = 0$. If we rewrite the modified Friedmann equations, given by Eqs. (\ref{friedmann}) and (\ref{acceleration}), in the standard form as in general relativity, we can define a torsion energy density, expressed by
\begin{equation}\label{rhoeff}
    \rho_{T} = -\frac{1}{16\pi G}[12H^2f'(T) + f(T)]\,,
\end{equation}
and a torsion pressure:
\begin{eqnarray}\label{peff}
    P_{T} &&= -\frac{1}{16\pi G}[48H^2f''(T)(\dot H+\frac{k}{a^2})\nonumber\\
    &&- 4f'(T)(3H^2+\dot H-\frac{k}{a^2}) - f(T)]\,,
\end{eqnarray}
both satisfying the \emph{torsion EoS}, $P_{T} = w_{T}\rho_{T}$ in which one defines the EoS due to torsion dark energy, dubbed $w_T$. It can be obtain by using Eqs. \eqref{peff} and (\ref{friedmann}), to give:
\begin{widetext}
\begin{eqnarray}\label{omegaeff}
    w_{T}&&=-1+\frac{12H^2f''(T)-f'(T)}{\frac{f(T)}{H^2}+12f'(T)}\times
[4\Omega_k\frac{1+f'(T)+12H^2f''(T)}{1+f'(T)-12H^2f''(T)}
    -\frac{6+12f'(T)+\frac{f(T)}{H^2}+6\Omega_k}{1+f'(T)-12H^2f''(T)}]+\nonumber\\ &&\frac{4\Omega_k(f'(T)+12H^2f''(T))}{\frac{f(T)}{H^2}+12f'(T)}\,.
\end{eqnarray}
\end{widetext}
We are interested in obtaining theoretical scenarios in which late-time cosmic speed up is driven by torsion only, without imposing any additional components in the Einstein energy momentum tensor. Due to these reasons, we take into account $f(T)$ models which have been calibrated with respect to the concordance paradigm\footnote{It is commonly accepted that viable alternatives to the cosmological constant $\Lambda$ provide as a limiting case at $z\simeq0$, the concordance paradigm. In other words, it seems that the best models capable of describing the universe dynamics today are those which reduce to a constant dark energy term at small redshift domains.}, i.e. the $\Lambda$CDM model, and we analyze their behaviors in the case of non-vanishing spatial curvature, through supernova data, combined with the information coming from baryon acoustic oscillation and the shift parameter of cosmic microwave background radiation. In particular, we consider here two $f(T)$ test functions which have been proposed in \cite{Wu1} for flat universe and we also match them with GRB observations.


\subsection{The First Model: polynomial torsional dark energy}

The first approach that we consider is a polynomial representation of $f(T)$. Bearing in mind that $T$ is negative-definite, one can simply imagine to take a certain polynomial order which dominates over matter. In particular, to check if the universe dynamics can be framed in the simplest hypothesis of polynomials, one can postulate a model depending upon two free constants only, under the form:
\begin{equation}\label{f}
f(T)=\alpha(-T)^n\,,
\end{equation}
with two free-parameters: $\alpha$ and $n$. Particularly, for given values of $n$ we get limiting cases which are given below:
\begin{eqnarray}
\left\{
  \begin{array}{ll}
    n=0, & \hbox{one recovers the $\Lambda$CDM case;} \nonumber\\
    n={1\over2}, & \hbox{one recovers the DGP model;} \nonumber\\
    n=1, & \hbox{one recovers standard cold dark matter model.}
  \end{array}
\right.
\end{eqnarray}
As stated above, we need our model to reproduce the concordance paradigm at small redshift domains. So that, our approach reduces to the $\Lambda$CDM model when $n=0$, including a limiting case which is compatible with observations at the very small redshifts. Further, it leads to the DGP model when $n={1\over 2}$, so that if observations indicate a significative matching with this number, the DGP model would degenerate with $f(T)$. The last reasonable case, i.e. the pressureless cold dark matter is valid if we re-scale Newton's constant. This value is however theoretically incompatible with dark energy, since it would decelerate the universe instead of pushing it up\footnote{This case is excluded by recent observations.}.

With those assumptions in mind, we leave unfixed both the coefficients. Hence, by using (\ref{friedmann}) and (\ref{f}), we rewrite the parameter $\alpha$ in terms of other cosmological parameters as
\begin{equation}\label{alpha}
\alpha=(6H_0^2)^{1-n}(1-\Omega_{k0})^{1-n}\frac{(1-\Omega_{m0}+\Omega_{k0})}{(2n-1+\Omega_{k0})}\,.
\end{equation}
Taking into account standard cosmological definitions, i.e. $\Omega_{m0}=8\pi G\rho/(3H^2)$ and $\Omega_{k0}=k/(a^2H^2)$, substituting the above expression into the modified Friedmann equations and defining $E^2 = H^2/H_0^2$, one gets
\begin{eqnarray}\label{ee}
    E^2&=&[(2n-1)E^2-\Omega_{k0}(1+z)^2]\left(\frac{1-\Omega_{m0}+\Omega_{k0}}{2n-1+\Omega_{k0}}\right)\times \\
&&\Big[\frac{E^2-\Omega_{k0}(1+z)^2}{1-\Omega_{k0}}\Big]^{n-1} +\Omega_{m0}(1+z)^3-\Omega_{k0}(1+z)^2\,.\nonumber
\end{eqnarray}


\subsection{The Second Model: Double torsional dark energy}

The second model is much more complicated than the first one and takes into account a phenomenological reconstruction of $f(T)$ which bids to two physical domains: the one with small torsion and the other with higher torsion values. In other words, this framework accounts for two different domains and candidates to be more complete than simple polynomials. In particular, defining $f(T)$ as follows:
\begin{equation}\label{f2}
    f(T)=-\alpha T\Big[1-\exp{\left(\frac{pT_0}{T}\right)}\Big]\,,
\end{equation}
we still have two free parameters, i.e. $\alpha$ and $p$, but we motivate $f(T)$ either because it leads to\footnote{We assume $T$ and $T_0$ negative and $p$ positive definite.}
\begin{eqnarray}
\left\{
  \begin{array}{ll}
    T\ll1, & \hbox{$f(T)\sim \exp{\left(\frac{pT_0}{T}\right)}$;} \\
    T\gg1, & \hbox{$f(T)\sim T$;} \\
  \end{array}
\right.
\end{eqnarray}
or in analogy with $f(R)$ models where an exponential dependence on the curvature scalar
is proposed, see e.g. for example \cite{Linder}, \cite{Bamba}. It is easy to check that when $p = 0$ one gets back the $\Lambda$CDM paradigm.

\noindent By using Eqs. (\ref{friedmann}) and (\ref{f2}), we can rewrite the parameter $\alpha$ in terms of other cosmological parameters as
\begin{equation}\label{alpha}
    \alpha=\frac{1-\Omega_{m0}+\Omega_{k0}}{2[1-e^p(1-p)]-(1-e^p)(1-\Omega_{k0})]}\,,
\end{equation}
and the modified Friedmann equations as
\begin{eqnarray}\label{ee}
    E^2&=&\Omega_{m0}(1+z)^3-\Omega_{k0}(1+z)^2-\\
    &&\alpha(E^2-\Omega_{k0}(1+z)^2)[1-\exp{(\frac{p(1-\Omega_{k0})}{E^2-\Omega_{k0}(1+z)^2})}]\nonumber \\ &+&2\alpha E^2[1-\exp{(\frac{p(1-\Omega_{k0})}{E^2-\Omega_{k0}(1+z)^2})}(1-\frac{p(1-\Omega_{k0})}{E^2-\Omega_{k0}(1+z)^2})]\,.\nonumber
\end{eqnarray}


\section{Cosmological Tools and experimental tests}

In this section, we examine the aforementioned models and we compare their evolutions with observational data, through the use of $\chi^2$ statistics and performing a numerical method developed on a grid. We thus investigate the constraints on the model parameters utilizing recent observational data surveys, including supernovae Ia which consist of 557 data points belonging to the Union 2.1 sample \cite{sne}, baryonic acoustic oscillation distance ratio and the shift parameter of cosmic microwave background radiation.

\noindent To do so, let us first recall the luminosity distance definition, as spatial curvature turns out to be non-zero. It reads \cite{Wu}:
\begin{equation}\label{dl}
D_{L}(z)\equiv\frac{(1+z)}{\sqrt{|\Omega_{k0}|}} {\cal{F}}\left(\sqrt{|\Omega_{k0}|}\int_0^z{\frac{dz'}{E(z')}}\right)\,,
\end{equation}
where ${\cal{F}}(x)\equiv(x, \sin(x), \sinh(x))$ for $k = (0, 1, -1)$.

The difference between the absolute and apparent luminosity of a distant object, i.e. the distance modulus, $\mu(z)$, is given by: $\mu(z) = 5\log_{10}D_L(z) - \mu_0$ where $\mu_0 = 5 log_{10}h - 42.38$ and $h = H_0/(100$km/s/Mpc). To constrain the parameters with supernovae, one employs the $\chi^2$ value:
\begin{equation}\label{chi2}
    \chi^2_{Sne}=\sum_{i=1}^{557}\frac{[\mu_i^{the}(z_i) - \mu_i^{obs}(z_i)]^2}{\sigma_i^2},
\end{equation}
where the sum is over the cosmological data points. Its minimum corresponds to maximizing the likelihood function $\propto \chi^3_{Sne}$ and gives us the best fit outcomes. In (\ref{chi2}), $\mu_i^{the}$ and $\mu_i^{obs}$ are the distance modulus obtained from the theoretical model and from observation respectively. Further, $\sigma_i$ is the estimated error of the $\mu_i^{obs}$, as obtained from observations. We notice that to best-fit by Union data, following \cite{Wu} we perform a marginalization on the present value of the distance modulus parameter $\mu_0$.

\begin{figure}[htbp]
\centering
\includegraphics[scale=0.3]{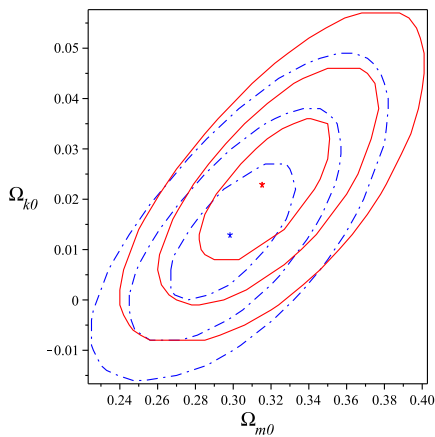}
\newline
\newline
\caption{\label{fig1} {\it{Contour plots and constraints on $\Omega_{k0}$ and $\Omega_{m0}$ at $1\sigma$, $2\sigma$ and $3\sigma$ confidence levels with the combinations of all data surveys. The solid red curves refer to the first model, whereas the dash-dot blue curves to the second one. The star in red and the cross blue point show best fit values.}}}
\end{figure}

For BAO data, the BAO distance ratio at $z = 0.20$ and $z = 0.35$ from the joint analysis of the 2dF Galaxy Redsihft Survey and SDSS data \cite{snss},\cite{snss2} is used. The distance ratio, given by
\begin{equation}\label{drbao}
   \frac{D_V(z=0.35)}{D_V(z=0.20)}=1.736\pm0.065,
\end{equation}
is a quasi model-independent quantity with $D_V(z)$ defined as
\begin{equation}\label{bao}
    D_V(z_{BAO})=\Big[\frac{z_{BAO}}{\Omega_{k0}H(z_{BAO})}{\cal{F}}^2(\sqrt{|\Omega_{k0}|}\int_0^{z_{BAO}}\frac{dz}{H(z)})\Big]^{\frac{1}{3}}.
\end{equation}
So, the constraint from BAO can be obtained by performing the following $\chi^2$ statistics
\begin{equation}\label{chibao}
    \chi^2_{BAO}=\frac{[(D_V(z=0.35)/D_V(z=0.20))-1.736]^2}{0.065^2}\cdot
\end{equation}
Finally, we add the cosmic microwave background shift in our analysis. Since the shift parameter $R$ \cite{r1}, \cite{r2}, contains the main information of the observations from the cosmic microwave background, it is used to constrain the theoretical models by minimizing
\begin{equation}\label{chicmb}
    \chi^2_{CMB}=\frac{[R-R_{obs}]^2}{\sigma_R^2},
\end{equation}
where $R_{obs} = 1.725\pm0.018$ \cite{r3}, is given by WMAP7 data. Its corresponding theoretical value is defined as
\begin{equation}\label{r}
    R\equiv\frac{\sqrt{\Omega_{m0}}}{\sqrt{|\Omega_{k0}|}}{\cal{F}}\left(\sqrt{|\Omega_{k0}|}\int_0^{z_{CMB}}\frac{dz}{E(z)}\right),
\end{equation}
with $z_{CMB} = 1091.3$. The constraints from a combination of all data surveys can be obtained by minimizing the net chi-squared $\chi_{tot}$, defined as:
\begin{eqnarray}\label{chitotale}
&&\chi_{tot}\equiv \sum_i\chi_{i}=\chi^2_{Sne}+\chi^2_{BAO}+\chi^2_{CMB}\,.
\end{eqnarray}
For the models that we proposed, the corresponding numerical outcomes have been reported in Tabs. \ref{table:1} and Fig. \ref{fig1}

\begin{table*}
\begin{center}
\caption{bestfit values of the first model} 
\centering 
\begin{tabular}{|c|c|c|c|c|} 
\hline\hline
Model Parameters  &  $\Omega_{m0}$ \ & $\Omega_{k0}$ \ & n and p \ & $\chi^2_{min}$ \\ [0.8ex] \hline
1st model & $0.315^{+0.019+0.038+0.057}_{-0.011-0.027-0.044}$ & $0.023^{+0.004+0.011+0.018}_{-0.008-0.015-0.023}$ & $-0.31^{+0.12+0.27+0.39}_{-0.16-0.41-0.70}$ &
543.486 \\ [2ex] \hline
2nd model & $0.298^{+0.014+0.033+0.053}_{-0.014-0.031-0.046}$ & $0.013^{+0.006+0.013+0.020}_{-0.005-0.013-0.020}$ & $0.16^{+0.12+0.33+0.57}_{-0.10-0.21-0.30}$ & 543.459\\ [2ex] \hline
\hline
\end{tabular}
{\tiny
\\
a. The parameters $n$ and $p$ refer to as the two models separately.

b. All parameters are not dimensional quantities.

}

\label{table:1}
\end{center}
\end{table*}\

\section{The use of gamma ray burst in $f(T)$ cosmology}

Data surveys of supernovae represent very accurate standard indicators. Their use, in particular, has been highly developed in the last decades. The disadvantage is that Union data are badly detectable at high redshifts. For example, current redshift limits are circumscribing around $z \simeq 2$. Thus, more distant regions cannot be investigated by using supernova data only. From these considerations, the need of more powerful far standard candles located at higher redshift domains is essential. To this end, the problem becomes particularly spinous at intermediate redshifts, i.e. when $z \simeq 6 \div 7$. Up to now, not very well-defined distance indicators are available.

\noindent This can be overcome, in part, by detecting GRBs, i.e. the most powerful universe explosions. In principle, their redshifts make these objects mostly appealing for possible uses in cosmology. Even though appealing, the use of GRBs cannot be considered at the same level of supernovae. Indeed, GRBs are not yet considered standard candles, since they do not provide known and well-defined luminosity-distance relations. Nevertheless, in several recent works, a wide number of detailed models aim at accounting either GRB formation or emission mechanisms, albeit none of them is intrinsically capable of connecting all the GRB observable quantities. However, there exist several observational correlations among some photometric and spectroscopic properties of GRBs. These correlations may enable to use them as distance indicators.

\noindent Actually, various attempts to build a GRBs Hubble diagram have yet been made. The author in \cite{Schaefer} has compiled a catalog of 69 GRBs with measured properties entering the five two parameters correlation then available. Even at the level of background cosmology, some authors have tried to compare the GRB diagram with cosmographic parameters, showing a suitable matching among numerical results obtained with type Ia supernovae than the ones got from GRBs. A recent development towards this topic has been recently discussed recently as Hubble's diagram, lately updated in \cite{Cardone}, has been reviewed by adding a new correlation. This new correlation is significative because can be used to either increase GRB's sample or to reduce uncertainties on $\mu(z)$.

\noindent In \cite{Cardone}, authors then re-calibrated all the six correlations considered using a fiducial $\Lambda$CDM cosmological model in agreement with the WMAP5 data and a Bayesian fitting technique. They built a new GRB Hubble diagram which depends on which cosmological background one employs. The technique itself provides also how to escape the circularity problem. Afterwards, in such a formalism it is possible to use Union 2.1 SNeIa sample in order to recover $\mu(z)$ in a model-independent way. Motivated by such approaches, we here test our models through (quite) model independent data samples as reported in \cite{Cardone}. In particular,

\begin{description}
  \item[1] A first data set of 83 GRBs. Hereafter, we  refer to it as \emph{fiducial GRB Hubble
diagram}.
  \item[2] Second, a data set of 69 GRBs. From now on, named the \emph{calibrated GRB Hubble diagram}.
\end{description}

\begin{figure}[htbp]
\centering
\includegraphics[width=0.40\textwidth]{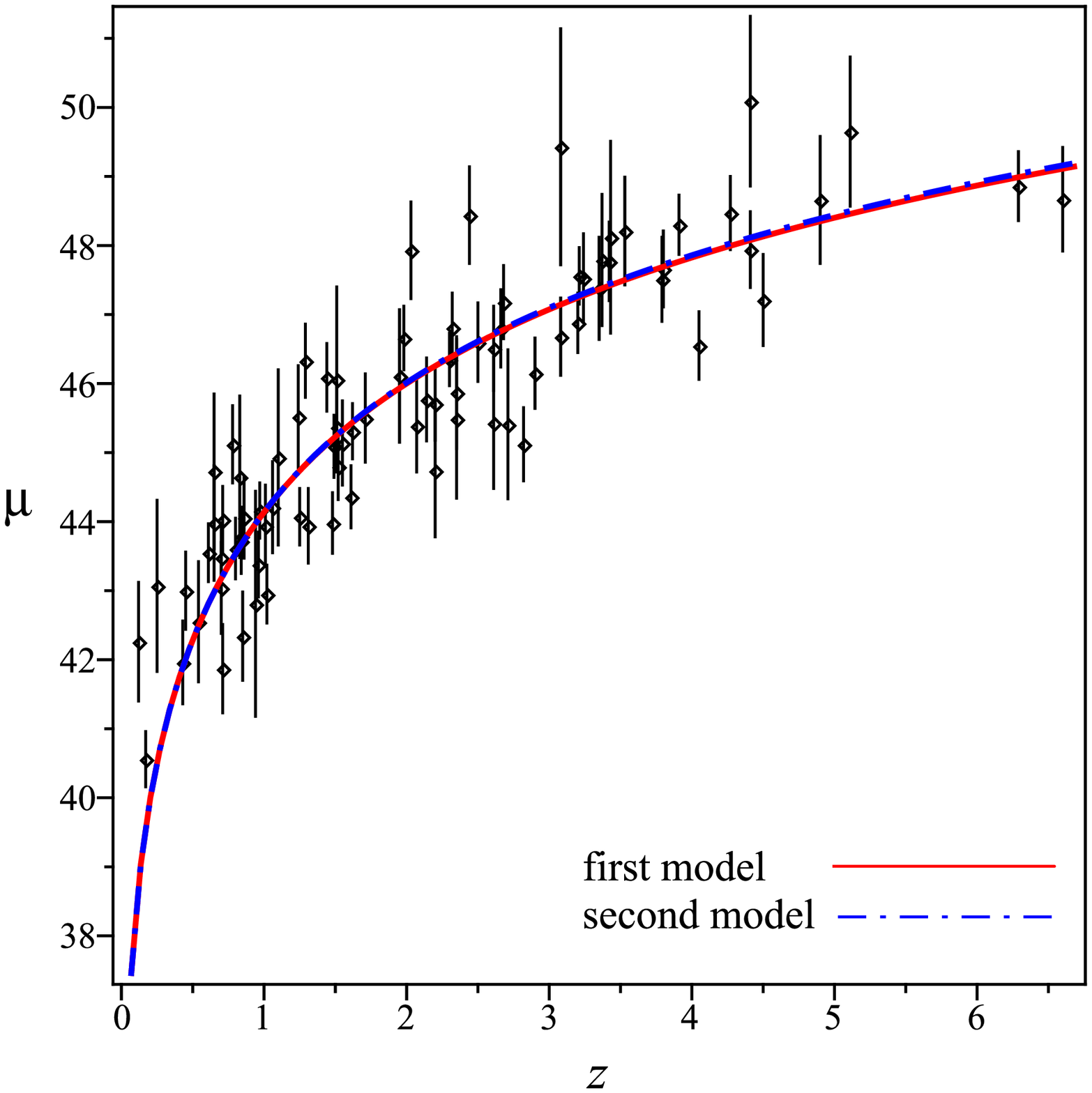}
\includegraphics[width=0.4\textwidth]{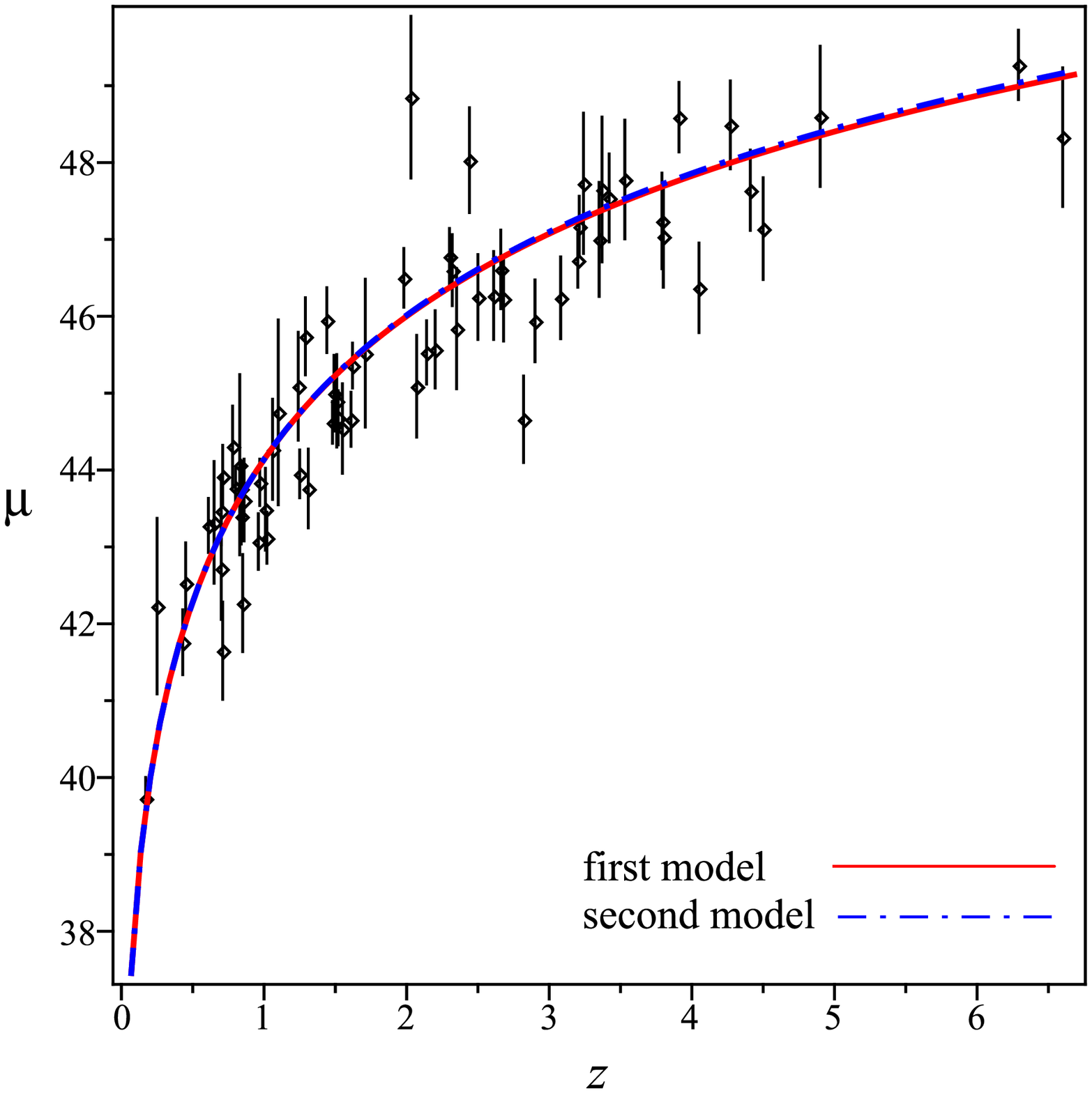}
\newline
\newline
\newline
\caption{{\it Left: The fiducial GRBs Hubble diagram with overplotted the distance modulus predicted by our both models. Right: The calibrated Hubble diagram with overplotted the distance modulus predicted by our both models.}}\label{fig2}
\end{figure}

The first data set is based on a calibration made by using a fiducial $\Lambda$CDM model to compute distances. The second survey, instead, has been built up with local regression-based methods. For all details, see Fig. \ref{fig2}.


\subsection{Discussion on numerical results}

Combined fits provide viable results with respect to modern observations since they are inside the Planck results \cite{planck}:
\begin{subequations}\label{equiv}
\begin{align}
\Omega_m&=0.315^{+0.016}_{-0.018}\,,\\
\Omega_\Lambda&=0.685^{+0.018}_{-0.016}\,,\\
\Omega_k&=-0.0005^{+0.0065}_{-0.0066}\,.
\end{align}
\end{subequations}
In particular, the value of $\Omega_{m0}$ is perfectly compatible with the above bounds, up to the 1$\sigma$ in both cases. The same can be concluded for spatial curvature. In this case, we notice that at the level of 1$\sigma$, spatial curvature is not compatible with zero. This is not true at higher $\sigma$, so that a definitive answer on the spatial geometry of the universe, in the context of $f(T)$ is not univocal. The $\chi^2$ are comparable between them, while the numerical bounds over the free coefficients $n$ and $p$ show that $n$ is negative and $p$ positive at 1$\sigma$. Unfortunately, due to the complexity of the second model, bounds over $p$ are much more plagued by higher error bars, at higher $\sigma$. In both cases, relative errors at the level of 3$\sigma$ are of the order of $\sim 0.5$, showing un-conclusive outcomes at the $\sim99\%$ of confidence level. However, the first model seems to be much more predictive since at 1$\sigma$, the relative errors are smaller than the second model. Cosmography, in the following discussion, can show whether at small redshift this is also confirmed analyzing the evolution of cosmographic coefficients around $z\simeq0$.

\section{Cosmography and torsional dark energy}

\begin{figure}[htbp]
\centering
\includegraphics[width=0.4\textwidth]{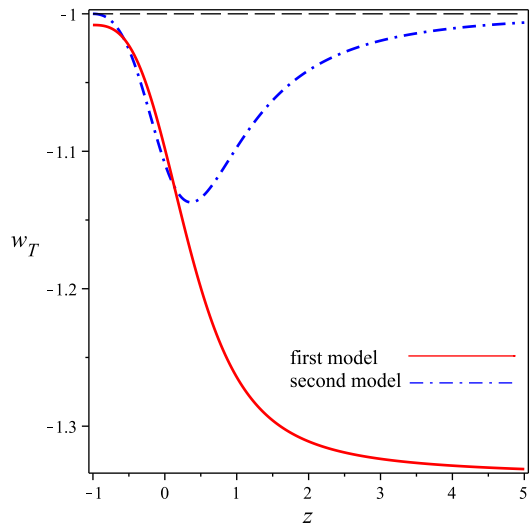}
\newline
\newline
\newline
\caption{{\it The evolutionary curves of the dark torsion EoS parameter for the best fitted values of our first (solid and red) and second model (dash-dot and blue).}}\label{fig4}
\end{figure}

Using the best-fitted model parameters, we can even discuss how torsion EoS behave in both the models. In Fig. (\ref{fig4}), we show the evolutionary curves of dark torsion EoS for the best fitted values of our models. One can see that the dark torsion EoS parameter becomes tangent to $w=-1$ as the redshift decreases for the  second model.

\subsection{The total EoS}

\begin{figure}
\centering
\includegraphics[width=0.4\textwidth]{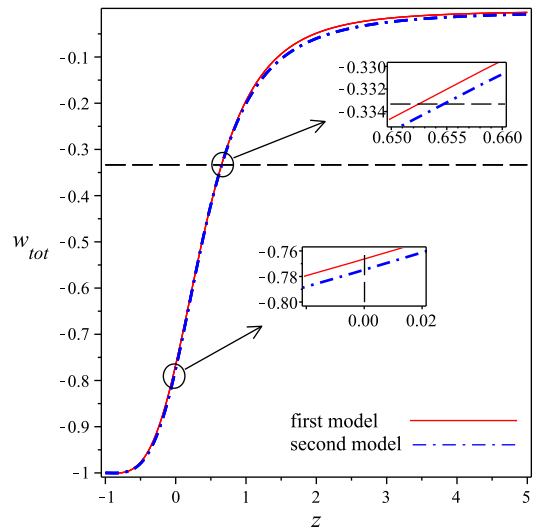}
\newline
\newline
\newline
\caption{{\it The trajectories correspond to the total EoS parameter as a function of $z$ for the best fitted values of our first (solid and red) and second model (dash-dot and blue).}}\label{fig5}
\end{figure}

The dynamics of the dark torsion EoS parameter has been discussed. Adding, not only torsion, but also cold dark matter and baryons implies correction terms, so that the total EoS parameter reads
\begin{equation}\label{omegatot}
    w_{tot}=-1+\frac{2}{3}\Big[\frac{(1+z)H^{-1}\frac{dH}{dz}+\Omega_{k}}{1+\Omega_{k}}\Big]\,.
\end{equation}
Using the best-fitted model parameters, obtained from $\chi^2$ method, one observes the evolution of the total EoS parameter $w_{tot}$ as a function of $z$ for both the models. To observe how $\omega_{tot}$ evolves in the near past and future, we can notice from Fig. (\ref{fig5}) that the universe transits from deceleration to acceleration eras when $w_{tot}=-1/3$ at about $z\sim0.652$ for our first model and about $z\sim0.655$ for our second model, approaching a de-Sitter phase in the future where $w_{tot}\rightarrow -1$. The change in the slope is known as transition redshift, i.e. the redshift at which the effective term of dark energy dominates over matter.

At $z=0$, we find $w_{tot}\simeq -0.77$ and $w_{tot}\simeq -0.78$ for the first and second models respectively. Unfortunately, those values both degenerate with the $\Lambda$CDM model. In the former case, in fact, one gets as   total EoS for the $\Lambda$CDM paradigm, the following:
\begin{equation}\label{wLtot}
w_{tot,\Lambda}=-\frac{1}{3}\frac{\Omega_k(z^2+2z-2)+3(1-\Omega_m)}{1+\Omega_kz(z+2)+\Omega_m(1+z)^3}\,,
\end{equation}
which provides as transition redshift
\begin{equation}\label{transLcdm}
z_{tr}=-\frac{\Omega_m + (-2)^{{1\over3}}\Big[\Omega_m^2\left(1 - \Omega_k - \Omega_m\right)\Big]^{{1\over3}}}{\Omega_m}\,,
\end{equation}
giving
\begin{equation}\label{results}
z_{tr}\simeq -0.685^{+0.020}_{-0.022}\,,
\end{equation}
with Planck results, where error propagation has been evaluated by using the standard logarithmic formula, i.e. $\delta z_{tr}=\sum_{i=m;k}\Big|\frac{\partial_i z}{\partial \theta_i}\Big|\delta\theta_i$, with $\theta_i\equiv\left\{\Omega_m,\Omega_k\right\}$.

The consequences of degeneracy on the onset of acceleration are that, from experimental bounds over $n$ and $p$, one concludes that $f(T)$ cosmology is compatible with the concordance paradigm, leaving open the possibility that our models are slight extensions of the $\Lambda$CDM paradigm. To figure this out, we can now approach the cosmographic method in order to distinguish if cosmographic coefficients depart from the fiducial case.


\subsection{Matching $f(T)$ models with cosmography of observable universe}

Any terms entering the total EoS naturally leads to a particular constituents inside the Friedmann equations. Postulating the dark energy EoS means that the thermodynamics of dark energy is known. It follows that the quest of understanding universe's expansion history is equivalent to postulate the EoS at different stages of universe's evolution. Finding out the most viable approximation to the $f(T)$ EoS determines a key towards understanding the micro-physics of the corresponding effective dark energy. Thus, one can wonder whether it is possible to reproduce the EoS in a model-independent manner.

\noindent In doing so, we do not require any specific cosmological models and so, assuming
$w = \sum_iP_i/ \sum_i\rho_i$, with the total pressure,
$P =\sum_iP_i$, and the total density,
$\rho = \sum_i\rho_i$, expanding into a Taylor series, we can predict the values of each derivatives by matching it with cosmic data. This strategy is known as \emph{cosmography}.

Cosmography aims at discriminating the class of models suitable for describing large scale dynamics at small redshift. So that, expanding the pressure, one gets
\begin{equation}
P = \sum_{k=0}^{\infty}\frac{1\,}{k!}\frac{d^{k}P}{dt^{k}}\Big|_{t_0}(t-t_0)^k
=\sum_{k=0}^{\infty}\frac{1\,}{k!} \frac{d^{k}P}{dy_i^{k}}\Big|_{0}y_i^k\,,
\end{equation}
where $y_i$ is an auxiliary variable of the form\footnote{The meaning of arbitrary new ``redshift'' variables has been introduced to overcome the convergence problem. It deals with the fact that most of cosmic data lie on redshift domains $z\geq1$, while Taylor expansions are built up at $z\simeq0$. Introducing $y_i$ would statistically favors the cosmographic analyses, instead of using $z$ only.}:

\begin{eqnarray}\label{hud}
y_i(z) &\rightarrow 1\,\,\,\,\,\,\,\,\, as\,\,\,\,\,\,\,\,\, z \rightarrow \infty\,,\nonumber\\
\,\nonumber\\
y_i(z) &\rightarrow 0\,\,\,\,\,\,\,\,\, as\,\,\,\,\,\,\,\,\, z \rightarrow 0\,.\nonumber
\end{eqnarray}
Easily, one gets:
\begin{subequations} \label{eq:pressureandD}
\begin{align}
 P &= \frac{1}{3}H^2 \left( 2q - 1\right)\,,\label{eq:pressure} \\
\frac{dP}{dt} & = \frac{2}{3} H^3 \left(1 - j\right)\,,
\end{align}
\end{subequations}
where we introduced the cosmographic set of parameters:
\begin{align}
&H\equiv \dfrac{1}{a}\dfrac{da}{dt} \ , \hspace{1cm} q\equiv -\dfrac{1}{aH^2}\dfrac{d^2a}{dt^2}  \label{eq:H&q} \\
&j \equiv \dfrac{1}{aH^3}\dfrac{d^3a}{dt^3}
\end{align}
named \textit{Hubble}, \textit{deceleration} and \textit{jerk} parameters, entering the scale factor expansion $a(t)=1+\sum_{k=1}^{\infty}\dfrac{1}{k!}\dfrac{d^k a}{dt^k}\bigg | _{t=t_0}(t-t_0)^k$. With simple algebra, combining Eqs. (\ref{eq:pressureandD}) and the definitions of EoS, we thus find:
\begin{equation}\label{eq:omega}
\omega_{tot} = \frac{2q-1}{3}\,.
\end{equation}
With our results, prompted in Tab. I, we soon infer:
\begin{equation}\label{sjs1000}
q_{0}^{(1)}=-1.017^{+0.025+0.075+0.151}_{-0.017-0.057-0.123}\,,
\end{equation}
and
\begin{equation}\label{sjs1001}
q_{0}^{(2)}=-0.633^{+0.062+0.221+0.487}_{-0.055-0.177-0.356}\,.
\end{equation}
Above, the error bars have been obtained by computing numerics from the standard logarithmic rule. In particular, we used:
\begin{equation}\label{error}
\delta q_0^{(i)}\equiv\sum_\kappa\Big|\frac{\partial q_0^{(i)}}{\partial \theta_\kappa}\Big|\delta \theta_\kappa\,,
\end{equation}
where $i=1;2$ refers to the first and second models respectively, reported as superscript in Eqs. \eqref{sjs1000} and   \eqref{sjs1001}, whereas $\kappa$ is associated to the typology of free parameters entering the deceleration parameter. So that, $\theta_\kappa\equiv\{\Omega_{m0},\Omega_{k0},n,p\}$, with $\delta\theta_\kappa$ the corresponding error. In the above cosmographic representation of $q_0^{(i)}$ we find that the second model seems to better match the concordance model.

\noindent This turns out to be compatible with our bounds coming from the statistical analyses performed in the previous section, because the first model well approximates a de-Sitter phase than the second. This implies that $q_0\simeq-1$, as we found. Moreover, both the approaches are clearly compatible with the standard $\Lambda$CDM paradigm, in which one has:
\begin{equation}\label{lcdmq0}
q_{0}^{\Lambda CDM}=-1 + \frac{1}{2}\left(2 \Omega_{k0}+3\Omega_{m0}\right)\,,
\end{equation}
which gives
\begin{equation}\label{lcdmq0}
q_{0}^{\Lambda CDM}=-0.528^{+0.034}_{-0.031}\,,
\end{equation}
using Planck data, at 1$\sigma$. This corresponds to small differences with respect to our two models. As already stated, the second model better fits the $\Lambda$CDM paradigm. Finally, we also get:
\begin{subequations}\label{basics}
\begin{align}
&\frac{dq}{dz}^{(1)}>0\,,\\
&\frac{dq}{dz}^{(2)}>0\,,\\
&j_0^{(1)}>0\,,\\
&j_0^{(2)}>0\,,
\end{align}
\end{subequations}
where we used again data coming from Tab. I.

\noindent The results over the variation of $q$ and $j_0$ indicate that the deceleration parameter changed signs in the past, as requested. Moreover our numerics certify the goodness at our time of both our approaches, even adding spatial curvature. Cosmography, in particular, seems to indicate that the second model is favored to mimic the $\Lambda$CDM paradigm. However, the result of $q$ for the first case indicates that the polynomial $f(T)$ model shows a de-Sitter phase, also compatible with current cosmic speed up. Summing up, since at small and higher redshift domains the two $f(T)$ choices appear to be compatible with the standard $\Lambda$CDM predictions, they figure as viable alternatives to frame the universe dynamics through an effective cosmological constant dark energy got from torsion, in non-flat homogeneous and isotropic universe.


\section{Final outlooks and perspectives}

In this work, we proposed how to generalize $f(T)$ models by means of non-vanishing spatial curvature $k$. In particular, we demonstrated that generalizing the models by adding a non-flat FRW universe gives non-trivial results at the level of the whole universe's dynamics. Indeed, it leads to refined constraints over the forms of either the torsion field or the free coefficients of cosmological $f(T)$ models. To figure this out, we analyzed two classes of models, i.e. the first a polynomial approach whereas the second a phenomenological scenario, built up in analogy to other approaches. In both the cases, we considered only those models which depend upon the fewest number of free coefficients possible. In particular, the two frameworks have been constructed through two free constants only, fixing one of those parameters by matter's value today. Afterwards, we matched those models with cosmic data, employing the recent data surveys of Union 2.1, baryonic acoustic oscillation and cosmic microwave background measurements. We emphasized that the two popular $f(T)$ models enable the crossing of the phantom divide line due to dark torsion.

\noindent We thus fixed the model parameters at 3-$\sigma$ confidence level and we showed that increasing the accuracy cannot allow one to remove at all the degeneracy with the concordance model, which is however contained in the aforementioned approaches. Once the numerical outcomes have been obtained, we reported their best fits and we discussed the bounds, showing the main differences between non-flat and flat $f(T)$ cosmology. We forecasted that there is in principle no need to fix \emph{a priori} a flat universe, since either open or close universes under the hypothesis of $f(T)$ gravity seem to be compatible with cosmic data, at least at the level of small redshift domains. Indeed, at 1$\sigma$ of confidence level, the value of $\Omega_{k0}$ is not bounded to zero, albeit it is compatible with Planck intervals. We employed GRBs to frame the behaviors of both the models at higher redshift intervals and we compare these results with the above ones. Finally, we matched the results with cosmography and we analyzed the cosmographic demands at the level of the EoS, deceleration and jerk parameters. We showed that the first model indicates a phenomenological de-Sitter phase while the second better adapts to the $\Lambda$CDM predictions.

\noindent Future developments will look at constraining the same models, with the hypothesis of non-flat cosmology, also at the level of early-time cosmology. We will also investigate the perturbation equations and power spectrum consequences at higher redshift domains.

\section*{Acknowledgment}
The work is partly supported by VEGA Grant No. 2/0009/16 and from COST Action CA15117  "Cosmology and Astrophysics Network for Theoretical Advances and Training Actions" (CANTATA), supported by COST (European Cooperation in Science and Technology). S. Capozziello and O. Luongo are supported by Istituto Nazionale di Fisica Nucleare (INFN). R. Pincak would like to thank the TH division at CERN for hospitality.

\end{document}